\pdfoutput=1
\documentclass[aps,twocolumn,floatfix,prb,showpacs]{revtex4}
\usepackage{graphicx,amsmath,bbm,color}

\newcommand{\ua}{\uparrow}
\newcommand{\da}{\downarrow}

\begin{document}

\title{Interaction effects on proximity-induced superconductivity\\ in semiconducting nanowires}
\date{\today}

\author{J.\ Danon}
\author{K.\ Flensberg}
\affiliation{Niels Bohr International Academy, and the Center for Quantum Devices, Niels Bohr Institute, University of Copenhagen, 2100 Copenhagen, Denmark}

\begin{abstract}
We investigate the effect of electron-electron interactions on proximity-induced $s$-wave superconductivity in one-dimensional nanowires. We treat the interactions on a self-consistent mean-field level, and find an analytic expression for the effective pairing potential in the presence of interactions, valid for a weakly tunnel coupled wire. We show that for a set of {\it two} nanowires placed in parallel on a superconducting substrate, the interaction-induced reduction of the pairing energy could result in the effective {\it inter}wire pairing potential exceeding the {\it intra}wire potential, which is one of the requirements for creating a time-reversal symmetric topological superconducting state in such a two-wire system.
\end{abstract}

\maketitle

\section{Introduction}

A topological superconductor has a gapped bulk spectrum and localized zero-energy Majorana modes at its edges and around vortices in the bulk.\cite{RevModPhys.83.1057} These Majorana states are quasiparticles with non-Abelian braiding statistics, and the prospect of using them for topologically protected quantum computing\cite{RevModPhys.80.1083} sparked an intense search for topological phases in condensed matter systems. A one-dimensional topological superconductor thus has localized Majorana fermions at its ends\cite{kitaev} and one could envision implementing braiding operations in a network of such wires.\cite{alicea:natphys} Concrete proposals how to create a one-dimensional topological superconducting state in a semiconductor-superconductor heterostructure\cite{PhysRevLett.105.077001,PhysRevLett.105.177002} were rapidly followed by pioneering experiments\cite{Mourik25052012,das:natphys,PhysRevB.87.241401} showing indeed signatures of localized Majorana end states.

All these proposals rely on inducing a spin polarization in the wire (i.e.~breaking of time-reversal symmetry), enabling the wire to mimic a spinless $p + ip$-wave superconductor. Interestingly, a few years ago it was realized that, in analogy to the quantum Hall and quantum spin Hall effects, also topological superconductivity can arise not only in systems with broken time-reversal symmetry, but also in time-reversal invariant systems.\cite{PhysRevLett.102.187001} A time-reversal symmetric topological superconductor has a gapped bulk and has Majorana modes emerging in (time-reversed) {\it pairs} at the edges of the sample. Despite the fact that a pair of Majorana fermions constitutes one regular fermion, it was argued that these time-reversed Majorana pairs might obey non-Abelian braiding statistics.\cite{liu:arxiv} Although their braiding statistics are not topologically protected,\cite{konrad:tri} there are situations in which they could be useful in the context of topological quantum computation.

Proposals for how to create a solid-state time-reversal symmetric topological superconductor included coupling of a single-band semiconductor to a more exotic type of superconductor, such as $d_{x^2-y^2}$-wave\cite{PhysRevB.86.184516} or $s_\pm$-wave.\cite{PhysRevLett.111.056402} Other suggestions were to use a conventional $s$-wave superconductor but couple it to a multi-layer interacting semiconductor,\cite{PhysRevLett.108.147003} to different topological insulators,\cite{loss:1}, or couple it with a phase difference to different parts of a single multi-band semiconductor.\cite{PhysRevLett.108.036803,PhysRevLett.111.116402} In essence, most of these proposals rely on having two pairs of time-reversed Fermi surfaces of which the effective pairing potentials have opposite sign.

Very recently, the proposed setup of a single two-channel nanowire (or equivalently two single-channel nanowires) coupled to a conventional $s$-wave superconductor was investigated in more theoretical detail.\cite{erikas:arxiv} It was found in Ref.~\onlinecite{erikas:arxiv} that the requirement of having opposite induced pairing potentials in the two channels (as was assumed in Ref.~\onlinecite{PhysRevLett.111.116402}) could be generalized to the condition $\Delta_{1}\Delta_2 < \Delta_{12}^2$, where $\Delta_{1(2)}$ is the induced pairing potential in channel 1(2) and $\Delta_{12}$ is an {\it interchannel} pairing potential, coupling electrons in one channel to holes in the other channel. For the case that the two channels live in different nanowires, it was suggested that repulsive electron-electron interactions could play an important role: They are expected to suppress the induced {\it intra}wire pairing,\cite{PhysRevLett.107.036801,PhysRevB.84.014503} whereas the {\it inter}wire pairing is barely affected due to very effective screening by the superconductor. If this suppression is strong enough, the system could thus enter a topological superconducting phase, even for $\Delta_1 = \Delta_2$. This result was supported by numerical Hartree-Fock and DMRG calculations, which indeed showed the emergence of a topological non-trivial phase in a two-channel nanowire coupled to an $s$-wave superconductor if strong enough electron-electron interactions were included.\cite{haim:tritops}

Electron-electron interactions can thus be a crucial ingredient for devising a time-reversal invariant topological superconductor, and a thorough understanding of their effect on the properties of proximity-coupled nanowires is quin\-tes\-sential. Bosonization,\cite{PhysRevLett.107.036801,PhysRevB.84.214528} DMRG,\cite{PhysRevB.84.014503} and numerical Hartree-Fock\cite{manolescu:arxiv} approaches indicated that interactions can indeed suppress the proximity-induced pairing in a nanowire (as well as enhance its spin-polarization, which is especially relevant for systems with broken time-reversal symmetry).

In this work we analyze in detail the effect of electron-electron interactions on the induced electron-hole pairing in a proximity coupled nanowire. Assuming a finite superconducting order parameter inside the wire, imposed by its contact with a nearby bulk superconductor, we use a self-consistent mean-field approach---basically the same approach as used for the numerical calculations in Ref.~\onlinecite{haim:tritops}---and show that we can arrive at analytic expressions for the interaction-induced reduction of the proximity-induced superconductivity in the wire. For the realistic case of weak tunnel coupling between the wire and the superconductor (i.e.~the tunneling rate $\Gamma$ into the superconductor smaller than the pairing energy $\Delta$ inside the superconductor) we find the suppression to be $(1+\upsilon \ln \frac{1}{\tau} )^{-1}$, where $\upsilon$ characterizes the ratio of the strength of the electron-electron interactions and the nanowire density of states, and $\tau$ is the small parameter $\tau \sim \Gamma/\Delta$. For large $\upsilon \ln \tfrac{1}{\tau}$ this suppression becomes strong, ultimately reducing the induced pairing to zero.

In the context of time-reversal symmetric topological superconductivity in a two-wire setup, one has to compare this suppressed pairing energy with the interwire pairing $\Delta_{12}$. This interwire pairing is assumed not to be affected by interactions, due to very effective screening by the superconductor. It is however expected to decay with increasing distance $d$ between the wires. For two identical nanowires placed in parallel on a superconducting substrate, we find this decay to be $\propto d^{-1/2}$ as long as $d$ is smaller than the coherence length of the superconductor. In that case, the requirement $\Delta_1\Delta_2 < \Delta_{12}^2$ translates into $k_{\rm F}^{\rm sc}d \lesssim (1 + \upsilon \ln \frac{1}{\tau})^2$, where $k_{\rm F}^{\rm sc}$ is the Fermi wave vector in the superconductor. This sets a clear boundary condition for experiments trying to realize time-reversal symmetric topological superconductivity in a two-wire system.

The rest of the paper is organized as follows. In Sec.~\ref{sec:model} we introduce the two-wire setup we have in mind, and we present the Hamiltonian with which we describe the electrons in the wires and the superconductor. In Sec.~\ref{sec:is} we first review how to understand the induced pairing in a single wire in terms of correlated tunneling of electrons and holes through the semiconductor-superconductor interface. Then we extend this description to a pair of wires and derive an effective interwire electron-hole pairing Hamiltonian as a function of the distance between the wires. We then include in Sec.~\ref{sec:int} electron-electron interactions in our description and reevaluate the {\it intra}wire pairing in the presence of interactions. In Sec.~\ref{sec:imp} we then place our results in the context of topological superconductivity in nanowire-superconductor hybrid systems.

\section{Model}\label{sec:model}

\begin{figure}[b]
\begin{center}
\includegraphics[width=82mm]{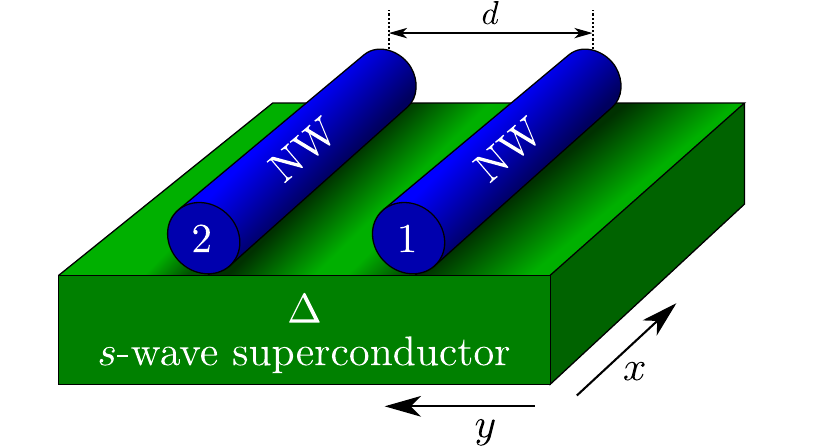}
\caption{(Color online) Model system investigated in this work: Two semiconducting nanowires are placed in parallel on a superconducting substrate. The electrons in the wires are tunnel coupled to the superconductor, and the superconductor has an $s$-wave pairing, its order parameter being $\Delta$.}
\label{fig:1}
\end{center}
\end{figure}
The system we have in mind is sketched in Fig.~\ref{fig:1}: Two semiconducting nanowires are deposited in parallel on a superconducting substrate, spaced by a distance $d$. Both wires are tunnel coupled to the same superconductor, which has an $s$-wave type pairing potential $\Delta$.

We assume both wires to be thin enough that only the lowest electronic subband is occupied. We write for the electrons in the wires the Hamiltonian
\begin{align}
\hat H_{{\rm nw},w} & = \frac{\hbar^2 k_{w}^2}{2m_{\rm nw}^*} - \mu_{\rm nw},\label{eq:hwire}
\end{align}
where $k_{w}$ represents the wave vector of the electrons in wire $w$, where $w \in \{1,2\}$, along the direction of the the nanowire axis, i.e.\ the $x$-axis. In this work we will exclusively focus on the interplay between the proximity effect due to the coupling to the superconductor and electron-electron interactions. Other ``standard'' ingredients, such a term describing spin-orbit interaction $\propto k_{w} \sigma_x$ and one describing a Zeeman field $\propto \sigma_z$ (where the Pauli matrices $\boldsymbol\sigma$ act in electron spin space), are disregarded for now.

The electrons in the $s$-wave superconducting substrate are described by
\begin{align}
\hat H_{\rm sc} = \frac{\hbar^2 p^2}{2m_{\rm sc}^*} - \mu_{\rm sc} - \Delta \sum_{\bf p} \left\{ \hat c_{{\bf p},\ua}^\dagger \hat c_{-{\bf p},\da}^\dagger + {\rm H.c.}
\right\},
\end{align}
$c_{{\bf p},\sigma}^{(\dagger)}$ being the electronic annihilation(creation) operator of an electron with momentum ${\bf p}$ and spin $\sigma$ in the superconductor. For convenience we take the pairing potential $\Delta$ to be real.

We assume the electronic states in the two nanowires to be tunnel coupled to those in the superconductor. We describe this coupling by the tunneling Hamiltonian
\begin{align}
\hat H_{{\rm t}} & = \sum_{w,\sigma} \int dx\, d{\bf r}'\,t_{w}(x,{\bf r}') \hat \psi_{w,\sigma}^\dagger (x) \hat \Psi_\sigma ({\bf r}') + {\rm H.c.}, \label{eq:ht}
\end{align}
where $\hat \psi_{w,\sigma}^\dagger (x)$ creates an electron with spin $\sigma$ at position $x$ in nanowire $w$, and $\hat \Psi_\sigma^\dagger ({\bf r}')$ creates an electron with spin $\sigma$ at position ${\bf r}'$ in the superconductor. The tunneling amplitudes are assumed to have the form
\begin{align}
t_w(x,{\bf r}') = \tilde t \,\delta(x'-x)\delta(y'-y_w)\delta(z'), \label{eq:tun}
\end{align}
where $y_w$ is the $y$ coordinate where wire $w$ touches the superconductor: Using $y_1 = 0$ and $y_2 = d$ corresponds with the geometry depicted in Fig.~\ref{fig:1}. For simplicity we assumed the coupling strength parameter $\tilde t$ (dimension energy times meters) to be equal for the two wires. Assuming translational invariance along the $x$-direction, we can Fourier transform the tunneling Hamiltonian (\ref{eq:ht}) to momentum space, which yields
\begin{align}
\hat H_{{\rm t}} & = \sum_{w,\sigma} \sum_{k,{\bf p}} t\delta_{k,p_x} \left\{ e^{-ip_yy_w} \hat a_{w,k,\sigma}^\dagger \hat c_{{\bf p},\sigma} + {\rm H.c.} \right\},
\end{align}
where $\hat a^\dagger_{w,k,\sigma} = L_x^{-1/2} \int dx\, e^{ikx}\hat\psi_{w,\sigma}(x)$ creates an electron with momentum $k$ and spin $\sigma$ in wire $w$, with $L_x$ being the length of the wires. We used the Kronecker delta function $\delta_{k,p_x}$ and we renormalized $t \equiv \tilde t\sqrt{L_x/ {\cal V}}$ with ${\cal V}$ the volume of the superconductor. The coefficient $t$ now has dimension energy, and for convenience it is assumed to be real.

Electron-electron interactions inside the nanowires are assumed to be very short-ranged due to strong screening by the nearby superconductor. We thus assume only intrawire interactions to be relevant and we model the interactions with a simple contact interaction potential, $U(r) = U\, \delta(r)$. This yields the Hamiltonian
\begin{align}
\hat H_{\rm ee} = \frac{U}{2L_x} \sum_{w,\sigma} \sum_{k,k',q} \hat a^\dagger_{w,k+q,\sigma}\hat a^\dagger_{w,k'-q,\bar\sigma}\hat a_{w,k',\bar\sigma}\hat a_{w,k,\sigma}. \label{eq:hee}
\end{align}
Note that only electrons with opposite spin interact, due to the Pauli principle.

We will treat the interactions on the mean-field level, such as was done in Ref.~\onlinecite{haim:tritops}. This gives us
\begin{align}
\hat H^{\rm mf}_{\rm ee} = \frac{U}{2L_x} & \sum_{w,\sigma}  \sum_{k,k'}\Big\{\langle \hat a^\dagger_{w,k',\bar \sigma}\hat a_{w,k',\bar\sigma}\rangle \hat a^\dagger_{w,k,\sigma}\hat a_{w,k,\sigma} \nonumber\\
& \quad + \langle \hat a^\dagger_{w,-k',\bar \sigma}\hat a^\dagger_{w,k',\sigma}\rangle \hat a_{w,k,\sigma}\hat a_{w,-k,\bar\sigma} \nonumber\\
& \quad + \langle \hat a_{w,k',\sigma}\hat a_{w,-k',\bar\sigma}\rangle \hat a^\dagger_{w,-k,\bar\sigma}\hat a^\dagger_{w,k,\sigma} \Big\},\label{eq:heemf}
\end{align}
where we used that our model is invariant under lattice translations along the $x$-direction and conserves spin. We stress again that our mean-field theory does not assume that the one-dimensional system itself is the source of breaking the gauge symmetry associated with superconductivity. The symmetry is already broken by the \textit{bulk} superconductor, and hence the order parameter $\langle \hat a^\dagger \hat a^\dagger \rangle$ is non-zero because of Cooper-pair tunneling from the bulk.

\section{Induced superconductivity: Non-interacting wires}\label{sec:is}

\subsection{One wire}\label{sec:peni}

We will now investigate the proximity-induced superconductivity in the nanowires. To make the paper self-contained, we first review the case of a single, non-interacting wire on a superconducting substrate, and show how one can understand the proximity effect in terms of correlated tunneling of electrons and holes through the semiconductor-superconductor interface.\cite{PhysRev.175.537} Our derivation closely follows similar derivations in existing literature, see e.g.\ Refs.\ \onlinecite{PhysRevB.82.094522} and \onlinecite{PhysRevB.84.144522}.

We start by integrating out the superconductor degrees of freedom, resulting in a self-energy term for the electron Green functions in the nanowire,
\begin{align}
\Sigma^{\rm sc} (k,i\omega_n) = t^2 \sum_{{\bf p}}\delta_{k,p_x}{\cal G}^{{\rm sc}} ({\bf p},i\omega_n),\label{eq:se2}
\end{align}
where $\omega_n = \pi (2n+1) T$ is a fermionic Matsubara frequency at temperature $T$ (setting $k_{\rm B} \to 1$). Both the self-energy $\Sigma^{\rm sc}$ and the superconductor's electronic Green function ${\cal G}^{{\rm sc}}$ are $2\times 2$ matrices in Nambu space. In terms of electronic creation and annihilation operators, we define ${\cal G}^{{\rm sc}} ({\bf p},i\omega_n)$ as the Fourier transform of the imaginary-time correlation functions,
\begin{align}
{\cal G}^{{\rm sc}} ({\bf p},i\omega_n) = \int_0^{1/T} d\tau\, e^{i\omega_n\tau} {\cal G}^{\rm sc} & ({\bf p},\tau),
\end{align}
with
\begin{align}
 {\cal G}^{\rm sc} & ({\bf p},\tau) = \nonumber\\
& \hspace{-.1em}-\left\langle \hat T_\tau \left(
\begin{array}{cc} 
\hat c_{{\bf p},\ua}(\tau)\hat c^\dagger_{{\bf p},\ua}(0)  &  \hat c_{{\bf p},\ua}(\tau)\hat c_{-{\bf p},\da}(0) \\
\hat c^\dagger_{-{\bf p},\da}(\tau)\hat c^\dagger_{{\bf p},\ua}(0)  &  \hat c^\dagger_{-{\bf p},\da}(\tau)\hat c_{-{\bf p},\da}(0)
\end{array}
\right)\right\rangle,
\end{align}
where $\hat T_\tau$ is the imaginary-time time-ordering operator. In this case, we find
\begin{align}
 {\cal G}^{\rm sc} ({\bf p},i\omega_n) = \frac{-1}{\omega_n^2 + (\varepsilon^{\rm sc}_{\bf p})^2 + |\Delta|^2} \left(
\begin{array}{cc} i\omega_n + \varepsilon^{\rm sc}_{\bf p} & -\Delta \\ -\Delta & i\omega_n - \varepsilon^{\rm sc}_{\bf p} \end{array}
\right),\label{eq:gsc}
\end{align}
where $\Delta$ is the order parameter of the ($s$-wave) superconductor, and the electron and hole energies $\varepsilon^{\rm sc}_{\bf p}$ are measured from the Fermi level, $\varepsilon^{\rm sc}_{\bf p} = \hbar^2p^2/2m_{\rm sc}^* - \mu_{\rm sc}$.

We see that the off-diagonal elements of the self-energy (\ref{eq:se2}) describe Andreev reflection at the semiconductor-superconductor interface, and thus introduce (superconducting) correlations between electrons and holes in the nanowire. We would like to describe this induced superconductivity with a simple electron-hole pairing term in the nanowire Hamiltonian, such as is usually done.\cite{PhysRevLett.100.096407,PhysRevLett.105.177002,PhysRevLett.105.077001} To this end, we convert the sum in (\ref{eq:se2}) into an integral,
\begin{align}
\Sigma^{\rm sc} (k,i\omega_n) = \frac{t^2 {\cal V}}{(2\pi)^2L_x} \int dp_y\, dp_z\, {\cal G}^{{\rm sc}} (k,p_y,p_z;i\omega_n).\label{eq:sesc}
\end{align}
Using that the Fermi energy of the superconductor $E_{\rm F}^{\rm sc}$ is typically much larger than its gap size $\Delta$, we assume that the normal-state density of states of the superconductor can safely be set to a constant. This allows us to perform the integral in (\ref{eq:sesc}), yielding
\begin{align}
\Sigma^{\rm sc} (k,i\omega_n) = \frac{\pi t^2 \nu_2}{\sqrt{\Delta^2 + \omega_n^2}} \left( \begin{array}{cc} -i\omega_n  & \Delta \\ \Delta & -i\omega_n \end{array} \right), \label{eq:sigmat}
\end{align}
where $\nu_2 \equiv (m_{\rm sc}^* / 2\pi \hbar^2)({\cal V}/L_x)$ is the effective two-dimensional normal-state density of states of the superconductor, which is probed by fixing $p_x = k$ and integrating over the other two momentum directions $p_{y,z}$.

The bare Green function of the electrons and holes in the nanowire reads
\begin{align}
{\cal G}^{(0)}(k,i\omega_n) = \frac{1}{i\omega_n -\tau_z \varepsilon^{\rm nw}_k},\label{eq:baregf}
\end{align}
where $\varepsilon^{\rm nw}_k = \hbar^2k^2/2m_{\rm nw}^* - \mu_{\rm nw}$ and $\tau_z$ is the third Pauli matrix in Nambu space, i.e.\ it yields $+1$ for electrons and $-1$ for holes. We then ``dress'' this Green function with the proximity self-energy (\ref{eq:sigmat}), which gives
\begin{align}
 {\cal G} (k,i\omega_n) = {\cal G}^{(0)} (k,i\omega_n) \big\{1- \Sigma^{\rm sc} (k,i\omega_n) {\cal G}^{(0)} (k,i\omega_n) \big\}^{-1}.
\end{align}
Analytic continuation of this Green function yields the retarded propagator
\begin{align}
 G^{\rm R} (k,\omega) = \frac{[1+\gamma(\omega)]^{-1}}{\omega - \displaystyle\frac{\varepsilon^{\rm nw}_k}{1+\gamma(\omega)}\tau_z - \frac{\gamma(\omega)\Delta}{1+\gamma(\omega)}\tau_x +i\eta},\label{eq:g1}
\end{align}
where $\eta = 0^+$ is a positive infinitesimal, and we introduced the frequency-dependent function $\gamma(\omega) = \pi t^2\nu_2 / \sqrt{\Delta^2-\omega^2}$. We see from (\ref{eq:g1}) that the tunnel coupling to the superconductor in general leads to (i) a reduced quasiparticle weight, (ii) a renormalization of the electron and hole energies, and (iii) a dynamical coupling between electrons and holes proportional to $\Delta$.

If the energies of the electrons of interest are much smaller than the gap size of the superconductor, $\omega \ll \Delta$, we can approximate $\gamma \approx \pi t^2\nu_2 / \Delta$, which is frequency-independent and characterizes the ratio between the tunneling rate into the superconductor and the size of the superconducting gap. In this case, the pairing term in the Green function (\ref{eq:g1}) can be equivalently produced by including the model Hamiltonian\cite{PhysRevB.84.144522}
\begin{align}
\hat H_{\rm pair} = \frac{\pi t^2 \nu_2 \Delta}{\pi t^2 \nu_2 + \Delta} \tau_x. \label{eq:pair}
\end{align}
We see that the magnitude of the effective induced pairing potential can range from zero to $\Delta$, depending on the ratio $\pi t^2\nu_2 / \Delta$. The regime of small $\pi t^2\nu_2 /\Delta$ is experimentally the most relevant\cite{PhysRevB.84.144522} and in this limit the induced pairing can be described by
\begin{align}
\hat H_{\rm pair} = \pi t^2 \nu_2\tau_x. \label{eq:pairr}
\end{align}

\subsection{Two wires}

If two separate wires are tunnel coupled to the same superconductor, as depicted in Fig.~\ref{fig:1}, we expect that besides the {\it intra}wire electron-hole correlations such as considered in the previous Section, also {\it inter}wire correlations could be induced. These correlations are the result of ``crossed Andreev reflection'' at the superconductor-semiconductor interfaces: A Cooper pair in the superconductor can scatter into a state with one extra electron in each wire, and vice versa.\cite{falci:europhys,PhysRevB.63.165314,loss:2,loss:3,PhysRevLett.111.060501}

To include such cross-wire processes in our description, the electron Green function and and self-energy of the nanowire part of the system are written as $4\times 4$ matrices in combined particle-hole and left-right nanowire space. The full self-energy then reads
\begin{align}
\Sigma^{\rm sc} (k,i\omega_n) = \left( \begin{array}{cc}
\Sigma^{{\rm sc},11}(k,i\omega_n) & \Sigma^{{\rm sc},12}(k,i\omega_n) \\
\Sigma^{{\rm sc},21}(k,i\omega_n) & \Sigma^{{\rm sc},22}(k,i\omega_n)
\end{array}\right),
\end{align}
in the two-nanowire space. The off-diagonal Nambu matrices are given by
\begin{align}
\Sigma^{{\rm sc},21}(k,i\omega_n) & = t^2 \sum_{{\bf p}}\delta_{k,p_x}e^{i p_y d} {\cal G}^{{\rm sc}} ({\bf p},i\omega_n),\label{eq:se3}
\end{align}
and $\Sigma^{{\rm sc},12}(k,i\omega_n)$ follows from substituting $d\to -d$. We see that these off-diagonal self-energies differ from (\ref{eq:se2}) by the factor $e^{\pm ip_yd}$ in the summand, taking care of the finite distance between the wires. The off-diagonal elements of $\Sigma^{{\rm sc},21}$ and $\Sigma^{{\rm sc},12}$ describe the aforementioned crossed Andreev reflection, whereas their diagonal elements describe virtual tunneling of electrons and holes between the two wires (via a quasiparticle state in the superconductor). 

A low-energy effective pairing Hamiltonian can be extracted in the same way as before. First, we convert the sum in (\ref{eq:se3})  into an integral over $p_{y,z}$ which yields in the same limit $E_{\rm F}^{\rm sc} \gg \Delta$ as we used before
\begin{align}
\Sigma^{{\rm sc},21}(k,i\omega_n) & = \frac{\pi t^2 \nu_2}{\sqrt{\Delta^2 + \omega_n^2}}\, g(d,\omega_n) \left( \begin{array}{cc} -i\omega_n  & \Delta \\ \Delta & -i\omega_n \end{array} \right),
\end{align}
where we defined the dimensionless function
\begin{align}
g(d,\omega_n) = \frac{2}{\pi} {\rm Im}\, K_0\left(d\sqrt{-(k^{\rm sc}_{\rm F})^2-i\tfrac{2m_{\rm sc}^*}{\hbar^2}\sqrt{\Delta^2+\omega_n^2}}\right),
\end{align}
with $K_0(x)$ being the zeroth order modified Bessel function of the second kind. In the limit of small energies, $\omega \ll \Delta$, we then straightforwardly find the effective interwire pairing term to be
\begin{align}
\hat H_{\rm pair}^{21} = \frac{\pi t^2 \nu_2\, g(d,0)\Delta}{\pi t^2 \nu_2\, g(d,0) + \Delta} \tau_x. \label{eq:pair12}
\end{align}

Let us now specialize this result to practically relevant conditions: (i) In a realistic setup, the interwire distance $d$ will be (much) larger than $(k^{\rm sc}_{\rm F})^{-1}$, so we use an asymptotic expression for the Bessel function\cite{Olver:2010:NHMF}
\begin{align}
\lim_{x\to\infty} K_0(x) = \sqrt{\frac{\pi}{2x}}e^{-x}.
\end{align}
(ii) We assume $\pi t^2 \nu_2 / \Delta$ to be small, which seems to be typically the case in experiments\cite{Mourik25052012,das:natphys,PhysRevB.87.241401} and is desirable when the goal is to create a topological superconducting state in the wire.\cite{PhysRevB.84.144522} Under these conditions we have
\begin{align}
\hat H_{\rm pair}^{21} = \pi t^2 \nu_2\sqrt{\frac{2}{\pi}} \frac{e^{-d/\xi_0} \sin( k^{\rm sc}_{\rm F}d + \tfrac{\pi}{4})}{\sqrt{k^{\rm sc}_{\rm F}d}}\tau_x, \label{eq:pair12r}
\end{align}
where the superconducting coherence length is defined as $\xi_0 = \hbar v_{\rm F}^{\rm sc} / \Delta$, with $v_{\rm F}^{\rm sc}$ being the Fermi velocity in the superconductor. It turns out that the approximation of the Bessel function is already excellent for moderately large $k^{\rm sc}_{\rm F}d$. From this Hamiltonian we thus extract the magnitude of the induced interwire pairing potential,
\begin{align}
\Delta_{12} =  \pi t^2 \nu_2\sqrt{\frac{2}{\pi}} \frac{e^{-d/\xi_0} \sin( k^{\rm sc}_{\rm F}d + \tfrac{\pi}{4})}{\sqrt{k^{\rm sc}_{\rm F}d}}.
\end{align}
We see that the pairing energy oscillates as a function of $d$ on a length scale $1/k^{\rm sc}_{\rm F}$, and is suppressed algebraically by $\propto (k^{\rm sc}_{\rm F}d)^{-1/2}$ and exponentially by $\propto e^{-d/\xi_0}$.

For two (zero-dimensional) quantum dots coupled to the same superconductor, the crossed Andreev amplitude was found to be proportional to $\propto (k^{\rm sc}_{\rm F}d)^{-1}e^{-d/\xi_0}$, where $d$ is now the distance between the dots.\cite{PhysRevLett.111.060501,PhysRevB.63.165314} We see that for the two-wire setup the suppression is less severe, we find for distances shorter than $\xi_0$ a $d^{-1/2}$-suppression instead of a $d^{-1}$-suppression. This ``gain'' of a factor $d^{1/2}$ is due to the reduced effective dimensionality of our setup: The translational invariance along the $x$-axis removes effectively one direction in which the electron-hole correlation function inside the superconductor decays. As a side remark, we note here that the suppression might become even less strong if one uses a diffusive superconductor instead of a clean one, as we assumed here.\cite{feinberg:epjb}

We also emphasize that the tunneling amplitudes used in our model (\ref{eq:tun}) consist of $\delta$-functions of coordinate. This implies that we assume that all tunneling between wire $w$ and the substrate takes place along a straight line with $y = y_w$ and $z = 0$. As soon as the wires have a finite contact area with the superconductor, and tunneling can take place within a band $d_y$ of $y$-coordinates around $y_w$, then the induced electron-hole correlations will be averaged accordingly. We see that the averaged pairing will be suppressed as soon as $k_{\rm F}^{\rm sc}d_y \gtrsim 1$ and drop to zero for large $k_{\rm F}^{\rm sc}d_y$. Since $(k_{\rm F}^{\rm sc})^{-1}$ is of the order of the lattice spacing in the superconductor, any realistic setup will be in the regime where $k_{\rm F}^{\rm sc}d_y \gtrsim 1$. However, as long as the suppression is not too severe, the order of magnitude of the interwire pairing term will still be $\sim t^2\nu_2 e^{-d/\xi_0}/\sqrt{k_{\rm F}^{\rm sc} d}$.

\section{Induced superconductivity: Interacting wires}\label{sec:int}

We now would like to include electron-electron interactions into our model, and see how their presence affects the superconducting correlations induced in the wires. To describe the interactions, we use the mean-field Hamiltonian (\ref{eq:heemf}) which includes only {\it intra}wire contact interaction, only between electrons with opposite spin. In this section we will thus focus again on a single wire.

The first (Hartree) term in Eq.~(\ref{eq:heemf}) leads to an energy shift of the electrons and holes that can be accounted for by an appropriate shift of the chemical potential. We will disregard this term and take the chemical potential as a fixed parameter, leaving us with
\begin{align}
\hat H^{\rm mf}_{\rm ee} = \frac{U}{2L_x} \sum_{k,k',\sigma}\Big\{ & \langle \hat a^\dagger_{-k',\bar \sigma}\hat a^\dagger_{k',\sigma}\rangle \hat a_{k,\sigma}\hat a_{-k,\bar\sigma} \nonumber\\
& + \langle \hat a_{k',\sigma}\hat a_{-k',\bar\sigma}\rangle \hat a^\dagger_{-k,\bar\sigma}\hat a^\dagger_{k,\sigma} \Big\}.\label{eq:heemf2}
\end{align}
The effect of this mean-field Hamiltonian is an extra term in the self-energy of the electrons and holes in the wire,
\begin{align}
\Sigma^{\rm tot} (k,i\omega_n) = \Sigma^{\rm sc} (k,i\omega_n) + \Sigma^{\rm int},\label{eq:sefull}
\end{align}
with
\begin{align}
\Sigma^{\rm int} = 
\frac{UT}{L_x} \sum_{q,m}\left(\begin{array}{cc}
0 & {\cal G}^{\rm eh} (q,i\omega_m) \\
{\cal G}^{\rm he} (q,i\omega_m) & 0
\end{array}\right),\label{eq:seint}
\end{align}
where $\omega_m = \pi(2m+1)T$ are again fermionic Matsubara frequencies. The functions ${\cal G}^{\rm eh}$ and ${\cal G}^{\rm he}$ are the off-diagonal elements of the electron Green function in Nambu space, explicitly reading
\begin{align}
{\cal G}^{{\rm he}} (k,i\omega_m) & = -\int_0^{1/T}\!\!\! d\tau\, e^{i\omega_m\tau}\langle \hat T_\tau \hat a^\dagger_{-k,\da}(\tau)\hat a^\dagger_{k,\ua}(0) \rangle, \\
{\cal G}^{{\rm eh}} (k,i\omega_m) & = -\int_0^{1/T}\!\!\! d\tau\, e^{i\omega_m\tau}\langle \hat T_\tau \hat a_{k,\ua}(\tau)\hat a_{-k,\da}(0) \rangle.
\end{align}
We note that the self-energy $\Sigma^{\rm int}$ does not depend on wave number or energy (resulting from our $\delta$-function approximation for the interaction potential). For convenience, we thus introduce the notation
\begin{align}
\Sigma^{\rm int} = 
\Delta
\left(\begin{array}{cc}
0 & s^{\rm eh} \\ s^{\rm he} & 0
\end{array}\right).\label{eq:ses}
\end{align}

The solution for the full Green function, dressed with the self-energy (\ref{eq:sefull}), can now be written as
\begin{align}
{\cal G}(k,i\omega_n) = {\cal G}^{(0)}(k,i\omega_n) \big\{1- \Sigma^{\rm tot}(k,i\omega_n) {\cal G}^{(0)}(k,i\omega_n) \big\}^{-1},\label{eq:full2}
\end{align}
and since $\Sigma^{\rm tot}(k,i\omega_n)$ depends via (\ref{eq:seint}) on ${\cal G}^{\rm eh}$ and ${\cal G}^{\rm he}$, this equation constitutes a self-consistency equation.

We insert the expression (\ref{eq:baregf}) for the bare Green functions into (\ref{eq:full2}) and write for the electron-hole correlation function ${\cal G}^{\rm eh}(k,i\omega_n)$
\begin{align}
{\cal G}^{\rm eh}(k,i\omega_n) = 
\frac{\displaystyle  -\Delta[ \gamma(i\omega_n) + s^{\rm eh}] }{\displaystyle \omega_n^2 + \frac{(\varepsilon^{\rm nw}_k)^2}{[1+\gamma(i\omega_n)]^2} + \frac{\Delta^2[\gamma(i\omega_n)+s^{\rm eh}]^2}{[1+\gamma(i\omega_n)]^2}}.\label{eq:geh}
\end{align}
Note that the argument of the function $\gamma(\omega)$ is now complex, which yields $\gamma(i\omega_n) = \pi t^2\nu_2 / \sqrt{|\Delta|^2+\omega_n^2}$. Due to our choice of real $t$ and $\Delta$, the analogous expression for the hole-electron correlation function ${\cal G}^{\rm he}(k,i\omega_n)$ is identical, and we see that $s^{\rm eh} = s^{\rm he} \equiv s$, with $s$ real.

Again specializing to the case of weak tunnel coupling, $\pi t^2 \nu_2/\Delta \ll 1$, which ensures that $\gamma(i\omega_n) \ll 1$, and summing (\ref{eq:geh}) over all allowed wave numbers $k$ and energies $\omega_n$ then yields a self-consistency equation for $s$,
\begin{align}
s = \frac{UT}{L_x} \sum_{q,n} \frac{-\gamma(i\omega_n) - s}{\omega_n^2 + (\varepsilon^{\rm nw}_q)^2+\Delta^2[\gamma(i\omega_n) + s]^2}. \label{eq:sc3}
\end{align}
In the standard way the sum over Matsubara frequencies is then rewritten as an integral enclosing the poles of a Fermi function and the sum over $q$ is converted into an integral. If we use the assumption $\tau \ll 1$, where $\tau\equiv \pi t^2\nu_2 / \Delta$, and furthermore require that $|\ln \tau|\gg 1$,\cite{note1} we can find approximate answers for the resulting integrals. At $T=0$, this finally yields the result
\begin{align}
s = -\upsilon(\tau+s) \ln \frac{1}{|\tau + s|},
\label{eq:s1}
\end{align}
where we introduced a dimensionless parameter characterizing the strength of the electron-electron interactions, $\upsilon \equiv U\nu_{\rm nw}(\mu_{\rm nw})/2\pi$ with $\nu_{\rm nw}(\mu_{\rm nw}) = (2m_{\rm nw}^*/\hbar^2\mu_{\rm nw})^{1/2}$ being the electronic density of states in the nanowire at the Fermi energy (i.e.~at the chemical potential, measured from the band edge where $q=0$). Alternatively, one can estimate the self-energy (\ref{eq:sc3}) by neglecting the frequency-dependence of $\gamma$. All integrals can then be performed analytically and, when assuming a constant density of states and using $\Delta$ as energy cut-off for electrons and holes contributing to Andreev reflection, the same result (\ref{eq:s1}) can in fact be produced.

We see from (\ref{eq:geh}) that we always must have $-\tau < s < 0$. For the case where $\tau$ is assumed to be small, $s$ is thus small as well. Under the present assumptions we can then use the approximate solution
\begin{align}
s = -\frac{\upsilon \tau \ln \frac{1}{\tau}}{1+\upsilon \ln \frac{1}{\tau}},\label{eq:sapp}
\end{align}
which depends on having large $|\ln\tau |$, but does not assume anything about $\upsilon$. Proceeding along the same lines as in Sec.~\ref{sec:peni}, we can write explicitly the retarded propagator $G^{\rm R}(k,\omega)$ for the electrons and holes, and then derive an effective pairing Hamiltonian which would provide the correct coupling between electrons and holes. Within our approximations, we find that $\hat H_{\rm pair} = \Delta_{1,2} \tau_x$, with
\begin{align}
\Delta_{1,2} = \frac{\pi t^2 \nu_2}{1+\upsilon \ln \frac{1}{\tau}},
\end{align}
i.e.~the pairing energy is a factor $(1+\upsilon \ln \frac{1}{\tau})$ smaller than the non-interacting potential found in Sec~\ref{sec:peni}. We see that for large $|\upsilon \ln \tau|$, the pairing energy becomes $\Delta_{1,2} = \pi t^2\nu_2 / \upsilon \ln \frac{1}{\tau}$, which ultimately tends to zero when $|\upsilon \ln \tau | \to \infty$. We again emphasize that this suppression plays a role only for {\it intra}wire induced superconductivity: The {\it inter}wire correlations, resulting from crossed Andreev reflection, are to first approximation not affected by the interactions, since they are assumed to be strongly screened by the superconductor.

\begin{figure}[t]
\begin{center}
\includegraphics[width=84mm]{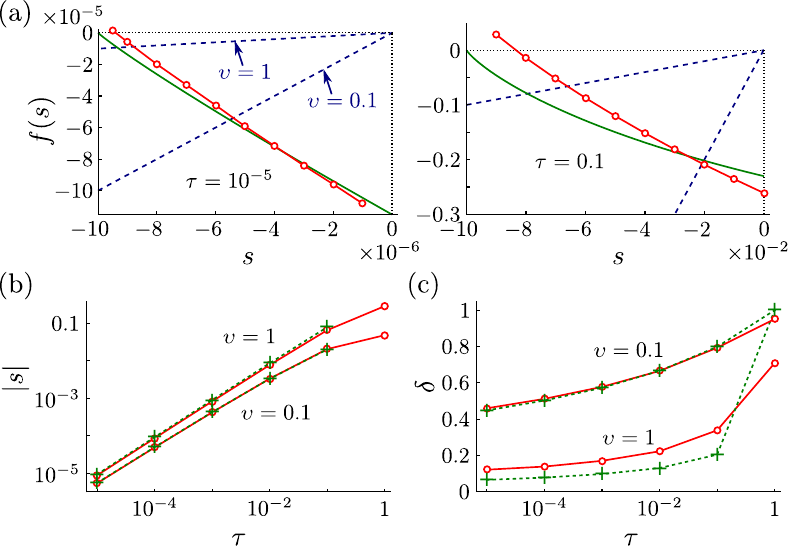}
\caption{(Color online) (a) Graphical presentation of numerical and analytic solutions to the self-consistent equation for $s$. The red connected dots are numerical evaluations of $(UT/\upsilon \Delta L_x)\sum_{q,m} {\cal G}^{\rm eh}(q,i\omega_m)$. The green solid lines show our analytic approximation of the same expression. The blue dashed lines show $s/\upsilon$ for $\upsilon = 1, 0.1$. Self-consistent solutions for $s$ occur at intersections of the blue dashed lines with either the numerical or analytic curves. In both plots we used $\mu_{\rm nw}/\Delta = 0.5$ and $T/\Delta = 10^{-6}$. (b) Values for $|s|$ found numerically (red circles) and with analytic approximations (green crosses). (c) Resulting relative suppression of $\Delta_{1,2}$.}
\label{fig:2}
\end{center}
\end{figure}
In the case of strong coupling, $\tau \sim 1$, Eq.~(\ref{eq:geh}) cannot be simplified to (\ref{eq:sc3}), and one has to find $s$ numerically from the self-consistency relation defined by (\ref{eq:seint}), (\ref{eq:ses}), and (\ref{eq:geh}). In Fig.~\ref{fig:2} we compare numerical results with our approximate expression (\ref{eq:sapp}). Fig.~\ref{fig:2}a shows graphical representations of the self-consistency relation, for $\tau = 10^{-5}$ and $\tau = 0.1$. In both plots we show three different functions of $s$ for the range of interest $-\tau < s < 0$: (i) The red connected dots present numerical evaluations of $(UT/\upsilon \Delta L_x)\sum_{q,m} {\cal G}^{\rm eh}(q,i\omega_m)$, with $\mu_{\rm nw}/\Delta = 0.5$ and $T/\Delta = 10^{-6}$. (ii) The blue dashed lines show $s/\upsilon$ as a function of $s$, for $\upsilon = 1$ and $\upsilon = 0.1$. (iii) The green solid lines show our small-$\tau$ expression $(\tau+s)\ln|\tau+s|$. Numerical self-consistent solutions for $s$ are given by the points where (i) and (ii) intersect, and solutions following from (\ref{eq:sapp}) correspond to points where (ii) and (iii) intersect. In Fig.~\ref{fig:2}b we show the values for $|s|$ found from the numerical calculation (red circles) and from the analytic approximation (green crosses). For $\tau = 1$, Eq.~(\ref{eq:sapp}) yields $s = 0$ for all $\upsilon$. Finally, Fig.~\ref{fig:2}c presents the resulting suppression of the intrawire pairing energy (close to the Fermi level), where we defined $\delta = \Delta_{1,2} (\upsilon) / \Delta_{1,2}(\upsilon = 0)$. We see that our small-$\tau$ approximation works reasonably well up to $\tau \sim 0.1$ and that it works better for small $\upsilon$.

\section{Implications and discussion}\label{sec:imp}

Let us now place our results in the context of the proposals for creating topological superconducting phases in nanowire-superconductor heterostructures. For topological superconductivity in single wires (with broken time-reversal symmetry), our results add to a detailed understanding of the physics of the proximity effect. In most theoretical descriptions, the proximity-induced superconductivity is incorporated in a phenomenological way, by introducing an effective electron-hole pairing term in the wire Hamiltonian, and electron-electron interactions are neglected. In this work, we arrived at analytical results showing the interplay between electron-electron interactions and the proximity effect for a one-dimensional semiconductor weakly coupled to a bulk superconductor.

For a {\it double}-wire setup however, our results are of even more importance. It has been shown that a set of two semiconducting nanowires coupled to the same $s$-wave superconductor can behave collectively as a time-reversal symmetric topological superconductor, one of the conditions being that $\Delta_{1}\Delta_{2} < \Delta_{12}^2$.\cite{erikas:arxiv} One of the proposed implementations used $\Delta_{12} = 0$, in which case it is required that the induced superconducting correlations in the two wires have opposite sign,\cite{PhysRevLett.111.116402} which poses an experimental challenge. As we showed above, assuming two identical wires placed in parallel on a superconducting substrate, one finds that (i) $\Delta_{1,2}$ are suppressed in the presence of interactions, whereas $\Delta_{12}$ can be assumed unaffected, and (ii) $\Delta_{12}$ becomes weaker with increasing distance $d$ between the wires, which to first approximation does not influence $\Delta_{1,2}$. This in principle allows for having $\Delta_{1},\Delta_{2} < \Delta_{12}$.

Combining our results from Secs.~\ref{sec:is} and \ref{sec:int}, we see that the criterion can be translated to
\begin{align}
\frac{e^{-d/\xi_0} \sin( k^{\rm sc}_{\rm F}d + \tfrac{\pi}{4})}{\sqrt{k^{\rm sc}_{\rm F}d/2\pi}} > \frac{\pi}{1+\upsilon \ln \frac{1}{\tau}}.
\end{align}
In the realistic regime where $(k_{\rm F}^{\rm sc})^{-1} < d \ll \xi_0$, we can reduce this inequality to
\begin{align}
k_{\rm F}^{\rm sc}d \lesssim (1 + \upsilon \ln \tfrac{1}{\tau})^2,
\end{align}
which sets a boundary on the interwire distance $d$. We see that both small tunnel coupling and strong interactions are favorable for satisfying the requirement.

For a quasi-one-dimensional nanowire we estimate
\begin{align*}
\upsilon \sim \frac{e^2}{4\pi^2\epsilon} \sqrt{\frac{2m^*_{\rm nw} }{\hbar^2 E_{\rm F}^{\rm nw}} },
\end{align*}
where $\epsilon = \epsilon_r\epsilon_0$ with $\epsilon_r$ the relative permittivity of the nanowire material. To arrive at quantitative estimates we assume common InAs nanowires, which have $\epsilon_r \approx 15$ and $m^*_{\rm nw} = 0.026m_e$. With $\mu_{\rm nw} \sim 0.1$~meV we then find $\upsilon \sim 2.5$.  Assuming that realistically $\ln \frac{1}{\tau}$ will not exceed 5 (but be more likely $\approx 2$), we find as a condition $k_{\rm F}^{\rm sc}d \lesssim 182$ (for $\upsilon \ln \frac{1}{\tau} = 12.5$) but more realistically $k_{\rm F}^{\rm sc}d \lesssim 36$ (for $\upsilon \ln \frac{1}{\tau} = 5$). With $k_{\rm F}^{\rm sc} \sim 10^{10}$ m$^{-1}$ this yields as maximum distance between the wires $\sim 20$ nm and $\sim 4$ nm for the two estimates.

Given that current-generation InAs nanowires have a diameter of 50--100 nm, this would most likely rule out their use for creating a two-wire-based time-reversal invariant topological superconducting state. In order to stretch the range of allowed distances, one could try to increase $\upsilon$ by working at a lower $\mu_{\rm nw}$, or with a different material with larger $m^*_{\rm nw}$ or smaller $\epsilon_r$. As noted before, using a diffusive superconductor might result in a less severe $d$-dependent suppression of the interwire pairing $\Delta_{12}$, which could also loosen the restrictions on $d$. Of course, another possibility is that the next generation of nanowire fabrication techniques can produce thinner wires: Our results indicate that a diameter of $\sim 10$ nm could be thin enough. Alternatively, one could imagine using instead of nanowires chains of adatoms placed directly on top of the superconducting substrate, such as was done in Ref.~\onlinecite{stevan:chain}: Then the two one-dimensional systems could in principle be atomically spaced.

Further, most proposals for nanowire-based topological superconductivity (both with or without time-reversal symmetry present) have significant spin-orbit coupling as a necessary ingredient for the emergence of a topological state. So, another relevant question to consider is how spin-orbit interaction would affect our results if we would incorporate it into our model. This would require adding a term $\alpha k_w \sigma_z$ to the wire Hamiltonian (\ref{eq:hwire}), where we chose the spin quantization axis along the direction of the effective spin-orbit field. This extra term can equivalently be produced by introducing a spin-dependent momentum shift, $k_w \to k_w + \sigma_z k_{\rm so}$, with $k_{\rm so} = \alpha m_{\rm nw}^* / \hbar^2$ and an appropriate shift of the chemical potential $\mu_{\rm nw}$. However, due to the contact interaction assumed in our model, the interaction self-energy (\ref{eq:seint}) contains a sum over all momenta and the momentum shift is thus irrelevant. If we furthermore take $\mu_{\rm nw}$ as measured from the band edge, then there is no difference in results at all.

Finally, one could question the accurateness of modeling the interaction energy as a contact interaction. A rough estimate for the actual screening length in the wires is given by their diameter. For wires with a diameter of $D \sim$ 50--100 nm, the corresponding energy scale $\varepsilon_{\rm scr} = \hbar^2/2m_{\rm nw}^*D^2 \sim 0.1$--0.6 meV is indeed not very large and might compete with other relevant energy scales in the system. As a very crude method to gain some insight in the effects of a finite screening length, one could investigate the self-energy $\Sigma^{\rm int}$ to first order in $U$, i.e.~evaluate Eq.~(\ref{eq:sc3}) with $s \to 0$ on the r.h.s. Assuming that $\gamma(i\omega_n) \equiv \tau$ is frequency-independent, we can perform the sum over Matsubara frequencies as well as the integral over momenta (using a constant density of states for the wire and assuming $\mu_{\rm nw} \sim \tau\Delta$). With a cut-off energy $\Delta$ and cut-off momentum $1/D$, we can evaluate the function $s(k)$ explicitly for small $k \ll 1/D$. We find in the small-$\tau$ limit
\begin{align}
s = -\upsilon\tau \ln\left( \frac{\zeta}{\tau} + \sqrt{\frac{\zeta^2}{\tau^2} + 1} \right),\label{eq:zeta}
\end{align}
with $\zeta = \varepsilon_{\rm scr}/\Delta$. We have per definition $\zeta < 1$ (otherwise all integrals would be cut off at $\Delta$). We thus see that as long as $\zeta \gg \tau$ (which corresponds to $\varepsilon_{\rm scr} \gg \tau\Delta$) the effect is negligible and (\ref{eq:zeta}) reduces to $s\approx -\upsilon\tau\ln\tfrac{1}{\tau}$, which agrees with our previous results to first order in $\upsilon$. For $\zeta \sim \tau$ the suppression of the self-energy is given by the full expression given above, and for $\zeta \ll \tau$, we find $s \approx -\upsilon\zeta$. We see that $\zeta \sim \tau$, where a finite screening length becomes important, corresponds to
\begin{align*}
\frac{\hbar^2}{2m_{\rm nw}^*D^2} \sim \tau\Delta.
\end{align*}
We always assume that $\mu_{\rm nw} \sim \tau\Delta$ so the regime where finite-range interactions can be neglected corresponds to $\mu_{\rm nw} \lesssim \hbar^2/2m_{\rm nw}^*D^2$. Our basic assumption that the wire can be considered quasi-one-dimensional thus automatically pushes us to the regime where the screening length can be safely set to zero.

\section{Conclusion}\label{sec:conc}

We have investigated in detail the effect of electron-electron interactions on the proximity-induced superconducting correlations in a one-dimensional nanowire. We treated the interactions in the wire on a self-consistent mean-field level, and found an analytic expression for the resulting effective pairing, valid for a weakly tunnel coupled wire ($\tau \ll 1$). Inspired by the theoretical proposal that a system of {\it two} nanowires coupled to the same $s$-wave superconductor can be driven into a time-reversal symmetric topologically non-trivial phase if the effective {\it inter}wire pairing $\Delta_{12}$ exceeds the {\it intra}wire pairings $\Delta_{1,2}$, we also derived an expression for the proximity-induced interwire pairing, which was found to decay with increasing interwire distance $\propto d^{-1/2}$. Combining these results, we translated the requirement $\Delta_1\Delta_2 < \Delta^2_{12}$ for creating a topologically non-trivial phase into a maximal distance between two parallel wires $k_{\rm F}^{\rm sc}d \lesssim (1 + \upsilon \ln \tfrac{1}{\tau})^2$, which sets a clear boundary for experiments.

We gratefully acknowledge very helpful discussions with Y.~Oreg, A.~Haim, E.~Gaidamauskas, K.~W\"{o}lms, and P.~Brouwer.


\begin{thebibliography}{10}

\bibitem{RevModPhys.83.1057}
X.-L. Qi and S.-C. Zhang, Rev. Mod. Phys. \textbf{83}, 1057 (2011).

\bibitem{RevModPhys.80.1083}
C.~Nayak, S.~H. Simon, A.~Stern, M.~Freedman, and S.~Das~Sarma, Rev. Mod. Phys.
  \textbf{80}, 1083 (2008).

\bibitem{kitaev}
A.~Kitaev, Phys. Usp. \textbf{44}, 131 (2001).

\bibitem{alicea:natphys}
J.~Alicea, Y.~Oreg, G.~Refael, F.~von Oppen, and M.~P.~A. Fisher, Nat. Phys.
  \textbf{7}, 412 (2011).

\bibitem{PhysRevLett.105.077001}
R.~M. Lutchyn, J.~D. Sau, and S.~Das~Sarma, Phys. Rev. Lett. \textbf{105},
  077001 (2010).

\bibitem{PhysRevLett.105.177002}
Y.~Oreg, G.~Refael, and F.~von Oppen, Phys. Rev. Lett. \textbf{105}, 177002
  (2010).

\bibitem{Mourik25052012}
V.~Mourik, K.~Zuo, S.~M. Frolov, S.~R. Plissard, E.~P. A.~M. Bakkers, and L.~P.
  Kouwenhoven, Science \textbf{336}, 1003 (2012).

\bibitem{das:natphys}
A.~Das, Y.~Ronen, Y.~Most, Y.~Oreg, M.~Heiblum, and H.~Shtrikman, Nat. Phys.
  \textbf{8}, 887 (2012).

\bibitem{PhysRevB.87.241401}
H.~O.~H. Churchill, V.~Fatemi, K.~Grove-Rasmussen, M.~T. Deng, P.~Caroff, H.~Q.
  Xu, and C.~M. Marcus, Phys. Rev. B \textbf{87}, 241401 (2013).

\bibitem{PhysRevLett.102.187001}
X.-L. Qi, T.~L. Hughes, S.~Raghu, and S.-C. Zhang, Phys. Rev. Lett.
  \textbf{102}, 187001 (2009).

\bibitem{liu:arxiv}
X.-J. Liu, C.~L.~M. Wong, and K.~T. Law, Phys. Rev. X \textbf{4}, 021018 (2014).

\bibitem{konrad:tri}
K. W\"{o}lms, A. Stern, and K. Flensberg, Phys. Rev. Lett. \textbf{113}, 246401 (2014).

\bibitem{PhysRevB.86.184516}
C.~L.~M. Wong and K.~T. Law, Phys. Rev. B \textbf{86}, 184516 (2012).

\bibitem{PhysRevLett.111.056402}
F.~Zhang, C.~L. Kane, and E.~J. Mele, Phys. Rev. Lett. \textbf{111}, 056402
  (2013).

\bibitem{PhysRevLett.108.147003}
S.~Nakosai, Y.~Tanaka, and N.~Nagaosa, Phys. Rev. Lett. \textbf{108}, 147003
  (2012).
  
\bibitem{loss:1}
J.~Klinovaja, A.~Yacoby, and D.~Loss, Phys. Rev. B \textbf{90}, 155447 (2014).

\bibitem{PhysRevLett.108.036803}
S.~Deng, L.~Viola, and G.~Ortiz, Phys. Rev. Lett. \textbf{108}, 036803 (2012).

\bibitem{PhysRevLett.111.116402}
A.~Keselman, L.~Fu, A.~Stern, and E.~Berg, Phys. Rev. Lett. \textbf{111},
  116402 (2013).

\bibitem{erikas:arxiv}
E.~Gaidamauskas, J.~Paaske, and K.~Flensberg, Phys. Rev. Lett. \textbf{112}, 126402 (2014).

\bibitem{PhysRevLett.107.036801}
S.~Gangadharaiah, B.~Braunecker, P.~Simon, and D.~Loss, Phys. Rev. Lett.
  \textbf{107}, 036801 (2011).

\bibitem{PhysRevB.84.014503}
E.~M. Stoudenmire, J.~Alicea, O.~A. Starykh, and M.~P. Fisher, Phys. Rev. B
  \textbf{84}, 014503 (2011).

\bibitem{haim:tritops}
A.~Haim, A.~Keselman, E.~Berg, and Y.~Oreg, Phys. Rev. B \textbf{89}, 220504(R) (2014).

\bibitem{PhysRevB.84.214528}
R.~M. Lutchyn and M.~P.~A. Fisher, Phys. Rev. B \textbf{84}, 214528 (2011).

\bibitem{manolescu:arxiv}
A.~Manolescu, D.~C. Marinescu, and T.~Stanescu, J. Phys.: Condens. Matter \textbf{26}, 172203 (2014).

\bibitem{PhysRev.175.537}
W.~L. McMillan, Phys. Rev. \textbf{175}, 537 (1968).

\bibitem{PhysRevB.82.094522}
J.~D. Sau, R.~M. Lutchyn, S.~Tewari, and S.~Das~Sarma, Phys. Rev. B
  \textbf{82}, 094522 (2010).

\bibitem{PhysRevB.84.144522}
T.~D. Stanescu, R.~M. Lutchyn, and S.~Das~Sarma, Phys. Rev. B \textbf{84},
  144522 (2011).

\bibitem{PhysRevLett.100.096407}
L.~Fu and C.~L. Kane, Phys. Rev. Lett. \textbf{100}, 096407 (2008).

\bibitem{falci:europhys}
G.~Falci, D.~Feinberg, and F.~W.~J. Hekking, Europhys. Lett. \textbf{54}, 255
  (2001).

\bibitem{PhysRevB.63.165314}
P.~Recher, E.~V. Sukhorukov, and D.~Loss, Phys. Rev. B \textbf{63}, 165314
  (2001).

\bibitem{loss:2}
P.~Recher and D.~Loss, Phys. Rev. B \textbf{65}, 165327 (2002).

\bibitem{loss:3}
K.~Sato, D.~Loss, and Y.~Tserkovnyak, Phys. Rev. B \textbf{85}, 235433 (2012).

\bibitem{PhysRevLett.111.060501}
M.~Leijnse and K.~Flensberg, Phys. Rev. Lett. \textbf{111}, 060501 (2013).

\bibitem{Olver:2010:NHMF}
F.~W.~J. Olver, D.~W. Lozier, R.~F. Boisvert, and C.~W. Clark, eds.,
  \emph{{NIST Handbook of Mathematical Functions}}, Cambridge University Press,
  New York, NY (2010).

\bibitem{feinberg:epjb}
D.~Feinberg, Eur. Phys. J. B \textbf{36}, 419 (2003).

\bibitem{note1}
The requirement $|\ln \tau| \gg 1$ is of course more severe than just $\tau \ll 1$, but this is the only regime in which analytic results could be produced. Results for $\tau \gtrsim 0.01$ should thus be interpreted in a more qualitative sense.

\bibitem{stevan:chain}
S. Nadj-Perge, I.~K. Drozdov, J. Li, H. Chen, S. Jeon, J. Seo, A.~H. MacDonald, B.~A. Bernevig, A. Yazdani, Science \textbf{346}, 602 (2014).


\end{thebibliography}
\end{document}